\documentclass[reprint,amsmath,amssymb,aps,preprintnumbers,nofootinbib]{revtex4-2}

\usepackage{graphicx}
\usepackage{dcolumn}
\usepackage{bm}
\usepackage{braket}
\usepackage[colorlinks=true, linkcolor=blue, citecolor=blue, urlcolor=blue]{hyperref}  
\usepackage{soul}  
\usepackage{enumitem}
\usepackage{color}
\usepackage{xcolor}
\usepackage{array}
\usepackage{float}
\usepackage{comment}
\usepackage[normalem]{ulem}

\newcommand{\equalcontrib}{Both authors contributed equally to this work.}
\newcommand{\axialhohfb}{{\sc AxialHOHFB }}

\newcolumntype{P}[1]{>{\centering\arraybackslash}p{#1}}

\begin{document}
\preprint{LLNL-JRNL-2022648}

\title{Microscopic Spin--Parity Distributions of Fission Fragments}

\author{Guillaume Scamps}
\thanks{\equalcontrib}
\affiliation{Universit\'e de Toulouse, CNRS/IN2P3, L2IT, Toulouse, France}

\author{Petar Marevi\'c}
\thanks{\equalcontrib}
\affiliation{University of Zagreb, Faculty of Science, 
Department of Physics, HR-10000 Zagreb, Croatia}

\author{Antonio Bjel\v{c}i\'c}
\affiliation{Nuclear data and Theory Group, Nuclear and 
Chemical Science Division, Lawrence Livermore National Laboratory, 
California, USA 94550}

\author{Nicolas Schunck}
\affiliation{Nuclear data and Theory Group, Nuclear 
and Chemical Science Division, Lawrence Livermore National Laboratory, 
California, USA 94550}

\begin{abstract}
Recent microscopic studies have investigated various 
features of the spin distributions of fission fragments, 
but their parity distributions
remain largely unexplored.
In this Letter, we provide a complete characterization 
of the spin--parity content of fission fragments within 
a time-dependent Hartree–Fock–Bogoliubov framework, 
performing for the first time simultaneous projections on angular momentum, 
particle number, and parity. 
Calculations are carried out for the thermal
neutron-induced fission of $^{239}$Pu using both 
the Gogny and Skyrme energy density functionals.
We find that dynamical pair breaking during fission populates a 
significant fraction of unnatural-parity states 
and generates components with non-zero spin projections $K$.
The parity content is found to depend on the number 
parity of the fragments, with odd-mass nuclei exhibiting 
pronounced parity staggering and odd-odd nuclei favoring 
negative parity.
These results show that the parity distribution of 
fragments can depart significantly 
from the equiprobable partition commonly assumed in 
statistical de-excitation models, 
with potential implications for the modeling of fragment decay.
\end{abstract}

\date{\today}

\maketitle

\textit{Introduction.} In nuclear fission, a single compound quantum system 
characterized by definite angular momentum and parity evolves into two 
separated fragments, each described by its own angular momentum and parity 
content~\cite{krappe2012,talou2023}.
Following the renewed interest sparked by recent experimental 
results~\cite{Wilson2021}, microscopic models \cite{schunck2022} 
have focused on determining 
the total angular momentum of the fragments, its
projection on the fission axis ($K$ quantum number), 
and correlations between the
fragments~\cite{bertsch2019,bulgac2021,marevic2021,scamps2023b,Marevic2026,li2026}.
However, these approaches do not fully characterize the combined 
angular momentum and parity content of the fragments, 
information that is essential to model their subsequent 
de-excitation and access their internal structure. Consequently, the population 
of parity states in the fragments -- both positive versus negative 
and natural versus unnatural -- remains unexplored within a microscopic framework.
In the absence of such predictions, state-of-the-art models of fragment decay, 
based on statistical reaction theory, 
typically assume a simple, equiprobable 
partition between positive and 
negative parities~\cite{talou2021,Kawano2023,verbeke2018,litaize2015}.

Parity is a fundamental symmetry of the nuclear wave function 
under spatial inversion, taking one of two values:
$\pi = +1$ (positive parity) or $\pi = -1$ (negative parity). 
This distinction, however, 
does not encode any relation to the angular momentum (spin)
$J$ of the state. 
A complementary classification is provided by 
the notions of natural and unnatural parity, 
which link parity directly to $J$.
States satisfying $\pi = (-1)^J$ (integer $J$) or $\pi = (-1)^{J-1/2}$ 
(half-integer $J$)
are referred to as natural parity states,
while those that do not are called unnatural parity states. 
In spherical shell-model configurations, 
natural parity states typically 
arise from normal single-particle excitations
within a given major shell, 
whereas unnatural parity states require cross-shell 
or more complex multi-quasiparticle excitations.
In odd-mass nuclei, the parity of the ground 
state and low-lying excitations is additionally
governed by the parity of the odd-particle
orbital. Consequently, unnatural-parity states can 
emerge directly from single-particle
degrees of freedom, without
invoking collective or multi-quasiparticle
excitations~\cite{bohr1970nuclear}.

From a reaction perspective, fission can be viewed as a quantum 
transition in which an initial compound state $(J_0,\pi_0)$ evolves 
into a two-fragment configuration characterized by $(J_1,\pi_1)$ and $(J_2,\pi_2)$, 
together with a relative orbital angular momentum $L$ carrying parity $\pi_L = (-1)^L$. 
Conservation of the total angular momentum
and parity imposes the constraints
$\bm{J}_0 = \bm{J}_1 + \bm{J}_2 + \bm{L}$ and $ \pi_0 = \pi_1 \, \pi_2 \, \pi_L$. 
The partitioning of these quantities
among the fragments can be understood in terms of the symmetry properties 
of the fragment wave functions and their breaking during the fission dynamics. 
Each broken symmetry leaves a characteristic fingerprint on the 
spin--parity distributions of the emerging fragments. 

For instance, a spherically-symmetric even-even fragment
would be characterized by a $J^\pi=0^+$ state.
Axial quadrupole deformations, which 
break rotational symmetry, generate even-$J$ configurations. 
Octupole correlations, associated with the breaking of reflection symmetry, 
induce negative-parity components and allow access to odd-$J$ configurations.
Further breaking of axial symmetry causes $K$ to no longer be a good quantum number, 
leading to non-zero $K$ components in the fragment spin distribution.
Such $K$-mixing can arise either
from a triaxial deformation of the fragment's
intrinsic density
or from a pair-breaking mechanism. In our approach,
as in most dynamical calculations
starting from an axial shape after saddle,
the fragments' shapes remain axially symmetric; the non-zero
$K$ components therefore originate entirely from the 
dynamical breaking of pairing correlations.
While a Cooper pair formed by two time-reversed 
nucleons carries $K=0 $ and positive parity, 
the violent dynamics at scission and the imparted 
excitation energy induce pair breaking.
This generates multi-quasiparticle configurations,
lifting the constraints of the paired vacuum
and naturally giving rise to non-zero $K$ and 
unnatural-parity components in the emerging fragments,
while the total $K_{tot}$ is exactly zero. 
The microscopic connection
between pair breaking, the resulting
non-zero fragment $K$ components,
and the emergence of unnatural-parity
configurations is discussed 
in the Supplemental Material \cite{SM}.

The goal of this work is to investigate these 
mechanisms microscopically, with particular emphasis on 
the previously unexplored role of parity and its 
connection to dynamical pair breaking. We study primary fragments 
produced in the thermal neutron-induced fission of $^{239}$Pu,
using two independent, state-of-the-art microscopic models 
based on the Gogny and Skyrme energy density functionals (EDFs).
Comparing their predictions
allows us to assess the model dependence of the predicted 
spin--parity distributions.

\textit{Method.} Fission is simulated using
the time-dependent Hartree-Fock-Bogoliubov 
(TDHFB) framework \cite{Bulgac2019b,simenel2025}
with both Gogny and Skyrme EDFs.
Gogny calculations are performed using
an in-house code
\cite{hashimoto2012,hashimoto2013,hashimoto2016,scamps2017,scamps2025}
in a hybrid basis
consisting of two-dimensional harmonic oscillator (HO)
eigenfunctions and a one-dimensional spatial mesh
along the fission axis. Skyrme calculations
are performed with the
\axialhohfb code
\cite{bjelcic2026}
in an optimized three-dimensional HO basis. 
In both cases, the time evolution of the
nuclear wave function is initiated from a static
configuration beyond the saddle point
in $^{240}$Pu, modeling the thermal neutron-induced
fission of $^{239}$Pu.
The fragments are followed up to the point
of full separation, defined
either as the point where their
centers of mass are separated
by more than $30$~fm (Skyrme)
or as the point where the distance between
their skins is $6$~fm (Gogny).

Both the Gogny and the Skyrme models employ very large bases
($N_{\rm{basis}} > 5700$ states)
and have recently been successfully applied
in fission studies \cite{scamps2025,bjelcic2026}.
However, it is worth noting that
there are some important differences between
the models, including the mean-field interaction
(Gogny D1S \cite{berger1984} vs Skyrme SkM* \cite{bartel1982}),
the pairing interaction
(Gogny vs mixed pairing with a quasiparticle cutoff 
\cite{dobaczewski2002}),
and the symmetry restrictions (\axialhohfb imposes
axial symmetry). 
Because the two calculations 
rely on independent computer 
codes with different underlying assumptions and different EDFs, 
their agreement provides a stringent test of the 
robustness of the predicted spin–parity mechanisms.

To extract the contributions of states with well-defined parities
for a given fragment ($\pi = \pm 1$) from the global, parity-breaking TDHFB 
wave function $\ket{\Psi}$, we introduce a localized parity projection operator. 
Let $\hat{\Pi}_F$ be the local space reflection operator acting 
exclusively on the single-particle coordinates within a spatial box defined 
around the fragment $F$ (where $F \in \{H, L\}$ for the heavy or light fragment, 
respectively), relative to the fragment's center of mass. The local parity 
projection operator $\hat{P}_F^{\pi}$ onto a subspace with a definite fragment 
parity is then defined as
\begin{equation}
\hat{P}_F^{\pi} = \frac{1}{2} \left( \hat{1} + \pi \hat{\Pi}_F \right),
\label{eq:parity_projector}
\end{equation}
where $\hat{1}$ is the identity operator. This operator satisfies the 
standard idempotency and orthogonality relations, namely $(\hat{P}_F^{\pi})^2 
= \hat{P}_F^{\pi}$ and $\hat{P}_F^{+} \hat{P}_F^{-} = 0$.

The probability of finding the fragment $F$ with a specific parity $\pi$ is 
then obtained by evaluating the expectation value of this localized projection 
operator with respect to the total wave function $\ket{\Psi}$ of the entire system.
When performed simultaneously with the local projections onto fragment particle 
numbers $N$ and $Z$, the total angular momentum $J$, and its intrinsic projection $K$, 
this allows us to determine the joint distribution
\begin{equation}
P_{F}(J,K,\pi,N,Z)
=
\langle \Psi |
\hat{P}^{N}_F
\hat{P}^{Z}_F
\hat{P}^{J}_{KK,F}
\hat{P}^{\pi}_F
| \Psi \rangle
\label{eq:joint_prob}
\end{equation}
for the fully separated primary fragments.
Here, $\hat{P}^{N,Z}_F$ and $\hat{P}^{J}_{K K, F}$
are, respectively, the operators projecting on good particle
numbers \cite{simenel2010,robledo2026} and angular momentum \cite{scamps2023b}
in the fragments. Together with $\hat{P}_F^\pi$, they form a set of
mutually commuting operators. 
The overlaps in \eqref{eq:joint_prob} are evaluated using the 
Pfaffian method \cite{robledo2009,bertsch2012,scamps2013,wimmer12}.
Note that both the angular momentum and parity projections
are performed in the fragment's frame of reference.
The distribution \eqref{eq:joint_prob} is normalized
to one in each fragment.
In the Supplemental Material we give
more details on the projection procedure
and show that a global particle-number projection
does not affect our conclusions \cite{SM}.

\begin{figure*}[t]
    \centering
    \includegraphics[width=0.99\linewidth]{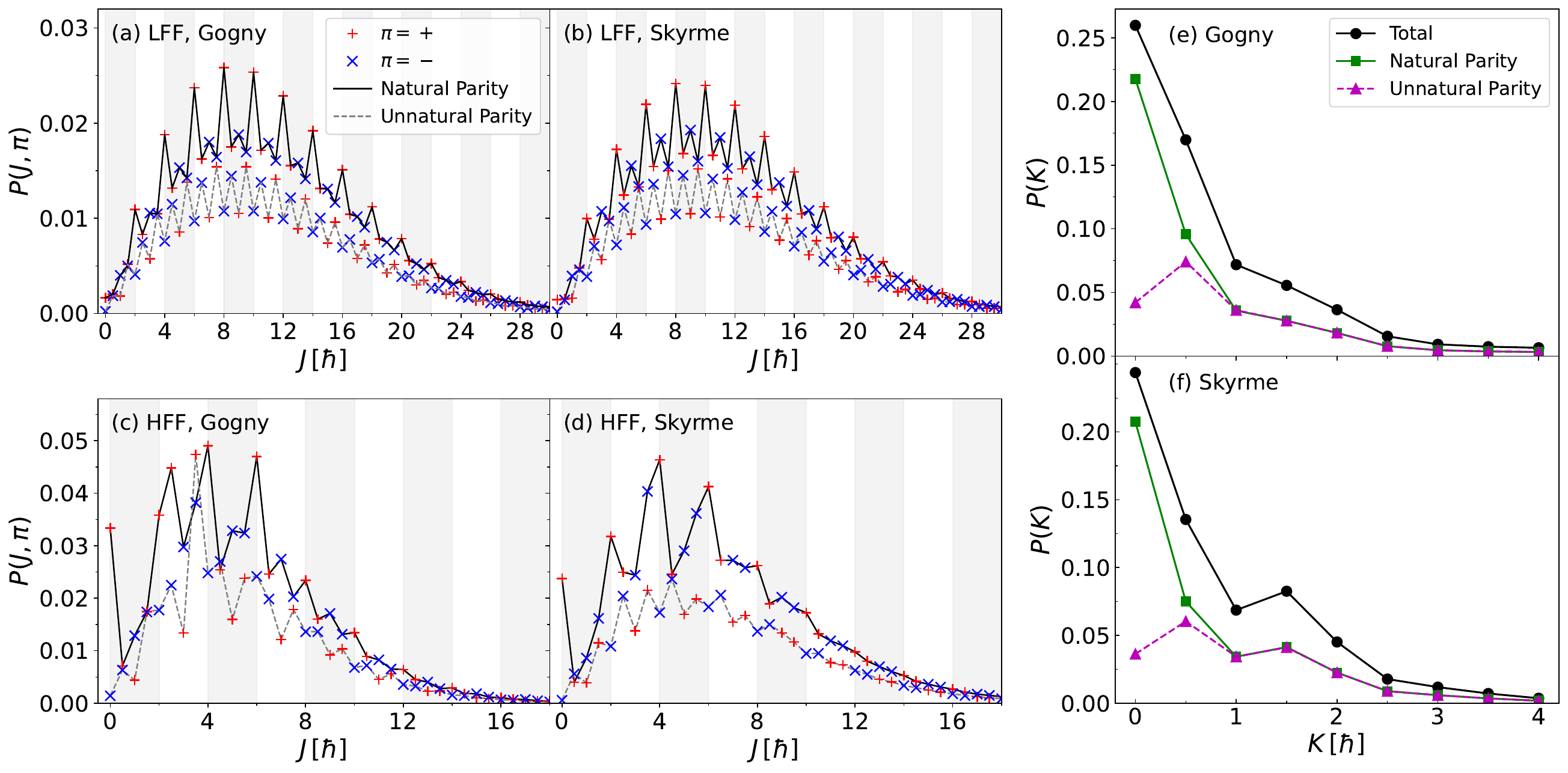}
    \caption{
    Panels (a)--(d): Spin--parity distributions $P(J,\pi)$ in 
    the light fission fragment (LFF) and
    the heavy fission fragment (HFF), calculated with Gogny and Skyrme EDFs. 
    Red pluses (blue crosses) indicate positive (negative) parity
    components. Black solid (gray dashed) lines connect states
    with natural
    (unnatural) parity.
    Alternating gray bands highlight even-$J$ intervals to guide the eye.
    Panels (e)--(f): Distribution $P(K)$ in the fragments,
    calculated with Gogny and Skyrme EDFs. Black circles
    correspond to the total distribution, while
    green squares (purple triangles) denote natural (unnatural)
    parity components.
    Note that the two fragments have the same distribution
    due to $K_{\rm{tot}} = 0$. Only $K \ge 0$ values are shown because 
    the distributions are symmetric around $K=0$. 
  }
  \label{fig:PJK_distributions}
\end{figure*}

\textit{Results.} In Fig.~\ref{fig:PJK_distributions}(a)--(d) we
show results for the primary fragments,
which are centered around $\braket{Z_H}=52.1$ 
and $\braket{N_H}\!=\! 82.5$ (Gogny), and 
$\braket{Z_H}=52.6$ and $\braket{N_H}=83.2 $ (Skyrme). 
The spin--parity distributions are 
obtained by marginalizing the full 
distribution \eqref{eq:joint_prob} over $N, Z$, and $K$.
Despite differences between the models,
Gogny [panels (a) and (c)] and Skyrme [(b) and (d)]
calculations yield very similar predictions. 
The spin distribution of the well-deformed ($\beta_2 \approx 0.50$) 
light fragment [(a) and (b)] is well described by a
standard spin-cutoff behavior, 
with an average value of $11-12\,\hbar$. Even-$J$,
positive-parity states clearly dominate, indicating only modest 
reflection-symmetry breaking at scission. Concurrently, 
natural-parity configurations comprise the vast majority of 
populated states at lower spins. 
However, in the odd-mass fragment channel, characterized by 
half-integer spins, this suppression 
is less pronounced and the population of unnatural-parity
states closely approaches that of natural-parity states.

In contrast, the nearly spherical  ($\beta_2 \approx 0.05$) heavy fragment [(c) and (d)] 
displays a significantly lower average 
spin of $5-6\,\hbar$, reflecting its more compact shape at scission. 
Its spin--parity landscape is more complex, exhibiting larger deviations from a simple 
spin-cutoff distribution. In particular, the $0^+$ state is prominently enhanced beyond 
statistical expectations. Furthermore, while natural parity generally dominates, this 
trend can occasionally be inverted. For instance, in Gogny calculations,
the $7/2^+$ state is populated with a 
noticeably higher probability than its natural $7/2^-$ counterpart.
Microscopically, these features can be interpreted as stemming from the
dynamical competition between preservation and breaking
of nucleon pairs.
The final fragment wave functions represent a quantum 
superposition of a fully paired vacuum state, 
which strictly preserves natural parity, 
and various multi-quasiparticle excitations. 
The non-negligible population of 
unnatural-parity states, alongside local state inversions, 
provides 
strong microscopic 
evidence of the crucial role of dynamical pair breaking during the rapid
transition to separated fragments.

To gain deeper insight into the interplay between the fragments' spin 
orientation and intrinsic structure, we analyze the $K$-dependence of 
the parity distributions. In Fig.~\ref{fig:PJK_distributions}(e)--(f), 
we show the  total $P(K)$ distribution in the fragments,
together with its decomposition
into natural- and unnatural-parity components.
Note that the fragments exhibit a non-trivial $P(K)$ distribution
despite being axially symmetric, since
$K$ is populated dynamically at scission
rather than fixed by the intrinsic shape \cite{scamps2023b}.
The two fragments share the same $P(K)$ distribution
because $K_{\rm{tot}} = 0$ and each fragment's
distribution is itself symmetric around $K = 0$.

The natural-parity states overwhelmingly dominate the $K=0$ mode, 
while the unnatural-parity contribution remains strongly suppressed. 
For the lowest non-zero projection ($|K|=1/2$), natural parity 
still maintains a noticeable, albeit less pronounced, dominance. 
However, a qualitative transition occurs as we move toward larger values 
of $K$ ($|K| \ge 1$). In this regime, the natural and unnatural 
parity curves converge, rendering the 
two contributions essentially equiprobable. 
This behavior reinforces the link between  
axial symmetry breaking, spin
alignment, and pairing dynamics. While the 
strongly paired $K=0$ base configurations strictly favor natural parity,
the multi-quasiparticle excitations required to build larger $|K|$ values 
naturally break time-reversal pairs. Consequently, the high-$|K|$ 
configurations carry an equal statistical mixture of natural and unnatural 
parity components, reflecting a pair-broken state.

\begin{figure*}[htbp]
    \centering
    \includegraphics[width=0.99\textwidth]{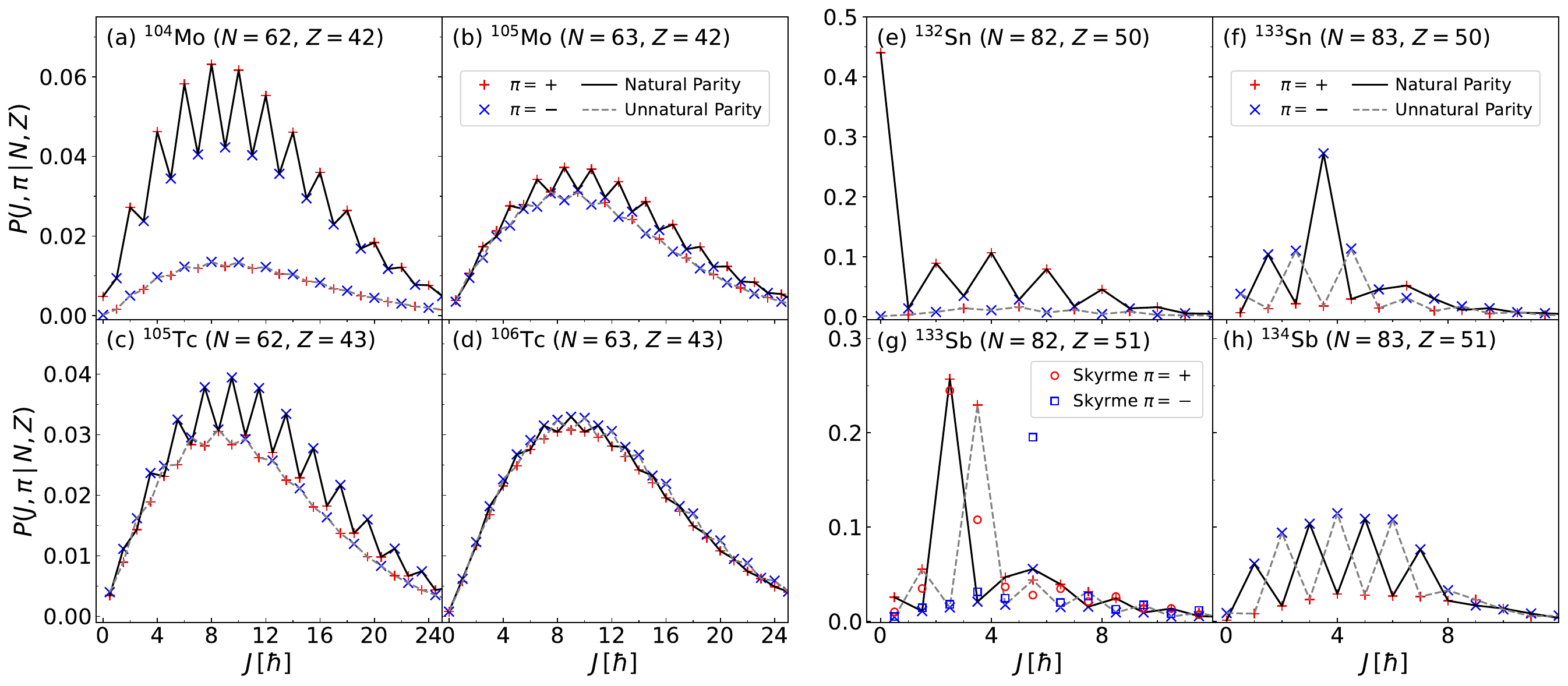}
    \caption{Spin--parity distributions $P(J, \pi\,|\,N, Z)$ of 
    several representative fragments with different neutron
    and proton number parities, calculated
    with the Gogny EDF.
    Panels (a)--(d) show light fragments, while panels
    (e)--(h) show heavy fragments.
     Red pluses (blue crosses) indicate positive (negative) parity
    components. Black solid (gray dashed) lines connect states
    with natural
    (unnatural) parity. As an example of the sensitivity 
    to the choice of EDF, the results for 
    $^{133}$Sb obtained with the Skyrme EDF are 
    shown in panel (g) for comparison.}
    \label{fig:pu240_fragments}
\end{figure*}

Due to the particle number projection, the distribution 
\eqref{eq:joint_prob} contains information
on a large number of fragments
around the most likely fragmentation. In the Gogny 
calculation, for instance, independent
projections in the two fragments yield
$218$ light
fragments and $162$ heavy fragments $(N_F, Z_F)$ whose weight in the total
distribution, $w(N_F,Z_F) = \sum_{J,K,\pi} P_F(J,K,\pi,N=N_F,Z=Z_F)$ 
exceeds $10^{-5}$. Note, however, that since these fragments
all originate from fluctuations around a single dynamical 
trajectory, they do not fully capture fragmentations 
associated with substantially different 
scission configurations, such as the symmetric 
or highly asymmetric split.

In the next step, we focus on properties of 
several representative fragments
with different neutron and proton number 
parities. In Fig.~\ref{fig:pu240_fragments},
we show the spin--parity distributions $P(J,\pi\,|\,N,Z)$ of
$^{104,105}\mathrm{Mo}$ ($Z=42$) and $^{105,106}$Tc ($Z=43$) for the light
fragment [panels (a)-(d)], and $^{132,133}$Sn ($Z=50$)
and $^{133,134}$Sb ($Z=51$) for the heavy fragment
[(e)-(h)]. These distributions
are obtained by fixing the number of nucleons ($N, Z$) 
to a given value and marginalizing 
the distribution \eqref{eq:joint_prob} over $K$
and $\pi$. All distributions have been renormalized to one, 
$\sum_{J,\pi}P(J,\pi\,|\,N,Z) = 1$.

The distributions exhibit a clear dependence on the 
neutron and proton number parity.
In the even-even $^{104}$Mo (a), natural-parity states dominate,
with an enhanced population of even-spin, positive-parity states. 
This behavior is consistent with a structure dominated by pairing 
correlations, where pair breaking is minimal and reflection symmetry 
is only weakly broken.
For odd-mass nuclides $^{105}$Mo (b) and $^{105}$Tc (c),
natural parity still dominates, but a marked staggering between 
positive and negative parity states is observed. This staggering 
appears to depend on whether the odd nucleon is a neutron or a proton.
In the present calculations,
odd-neutron fragments are found to favor positive-parity states, 
while odd-proton fragments favor negative-parity states. 
For the odd neutron around $N \simeq 62$,
this reflects the predominantly positive-parity character 
of the $sdg$ ($N_{\rm{sh}} = 4$) shell. 
For the odd proton around $Z \simeq 42$, 
the situation is more delicate: at the large deformations 
reached by the light fragment, the positive-parity 
$g_{9/2}$ orbitals and the negative-parity $pf$ 
($N_{\rm{sh}} = 3$) orbitals lie close together near the Fermi surface, 
and the calculations 
favor negative-parity configurations.
Furthermore, the odd-odd nucleus $^{106}$Tc (d)
exhibits a slight dominance of negative-parity states. 
This can be understood as a consequence of coupling 
an odd proton and an odd neutron
with opposite intrinsic parities, which increases 
the probability of producing negative-parity configurations 
in the final fragment state distribution.

\begin{figure*}[htbp]
    \centering
   \includegraphics[width=0.99\textwidth]{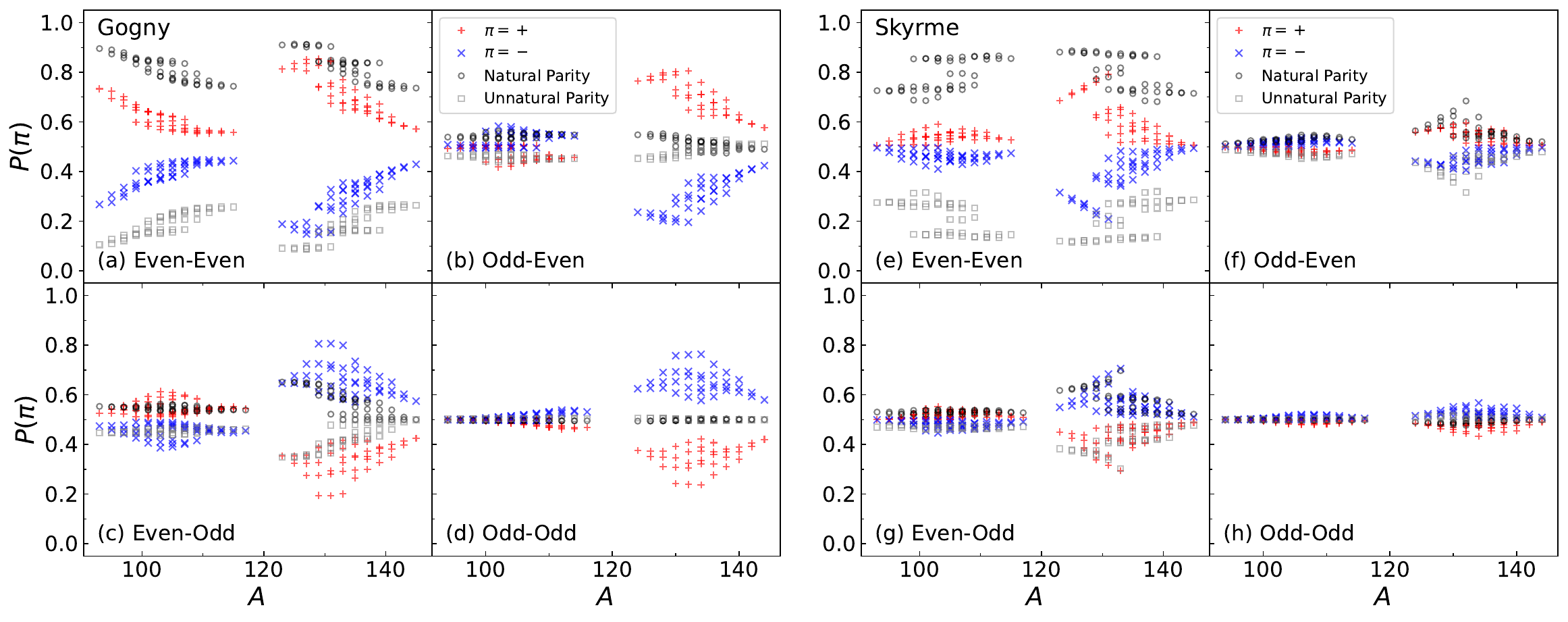}
   \caption{
   Parity distributions as a function of fragment mass number $A$, 
   partitioned by fragment category: 
   Even-Even ($Z$-even, $N$-even), Odd-Even ($Z$-odd, $N$-even), 
   Even-Odd ($Z$-even, $N$-odd), and Odd-Odd ($Z$-odd, $N$-odd). 
   For each fragment $(N,Z)$, the fraction of the yield carrying 
   positive (red pluses), negative (blue crosses), natural (open black circles), 
   and unnatural (open gray squares) parity is shown. 
   Only fragments with $N \in [77, 88]$, $Z \in [46, 58]$ (heavy) and 
   $N \in [57, 69]$, $Z \in [36, 48]$ (light) are included. 
   Results are obtained with the Gogny (left) and Skyrme (right) EDFs.}
   \label{fig:parity_vs_mass}
\end{figure*}

The situation is somewhat different for the heavy fragments, 
located in the vicinity of the double shell closure. 
In $^{132}$Sn (e), the $0^+$ state overwhelmingly dominates 
the distribution, reflecting the nearly spherical character 
of this doubly-magic nucleus. The remaining population is 
concentrated in low-$J$ states, predominantly of natural parity, 
with a clear preference for even-$J$ values. 
Adding one proton to the doubly-magic core leads to a 
spin distribution dominated by the $5/2^+$ and $7/2^+$ states
in $^{133}$Sb (g). 
In a simple shell-model picture, the first proton orbitals 
above the $Z=50$ shell closure are the $2d_{5/2}$ and 
$1g_{7/2}$ states, both carrying positive parity~\cite{ring2004}. 
Similarly, adding one neutron produces a distribution dominated 
by the $7/2^-$ state in $^{133}$Sn (f),
which can be associated with the 
occupation of the $2f_{7/2}$ orbital located above 
the $N=82$ shell gap.

These results suggest a remarkably simple interpretation 
of spin distributions in this region. The heavy fragment can be 
viewed as a doubly magic $^{132}$Sn core, predominantly in 
its $0^+$ ground state, coupled to an additional valence nucleon 
occupying a low-lying single-particle orbital. The resulting 
fragment spin then directly reflects the spin
and 
parity of this valence particle. Such a behavior illustrates 
the strong influence of shell structure on the quantum states 
populated in the fission fragments and demonstrates how simultaneous 
projections on particle number, angular momentum, and parity
can serve as a proxy to access this underlying microscopic structure.

Although the Gogny and Skyrme EDFs predict the same 
microscopic mechanisms governing the spin--parity distributions, 
they can nonetheless
produce different single-particle structures in the fragments. 
This is illustrated in Fig.~\ref{fig:pu240_fragments}(g)
for the $^{133}$Sb ($Z=51$) fragment.
In both calculations, 
the $5/2^+$ state is the most strongly populated. However, 
the Skyrme functional also predicts a significant population of 
the $11/2^-$ state, reflecting the influence of the $1h_{11/2}$ 
orbital in the underlying single-particle spectrum.

To assess the global impact of the underlying 
single-particle structure on the parity distributions, 
Fig.~\ref{fig:parity_vs_mass} shows the probabilities of 
positive- versus negative-parity, and natural- versus unnatural-parity, 
configurations as a function of fragment mass.
Despite some quantitative differences,
the two EDFs provide a consistent qualitative
picture.
The largest deviation from equiprobable
partition is observed for even-even
fragments, where the probability of preserving 
paired configurations is highest. For both
the heavy and the light even-even fragments,
positive and natural parity strongly dominate
at low excitation energies, in contrast to 
the assumptions commonly adopted in current fission-fragment decay models.

For odd-even, even-odd, and odd-odd fragments,
the deviation from equiprobable partition
is much more pronounced in heavy fragments.
For odd-even and even-odd fragments, in particular,
the difference between natural 
and unnatural parity is rather modest.
In these cases, 
the parity is largely determined by the unpaired nucleon. 
For the heavy proton-odd and light neutron-odd fragments, 
the unpaired nucleon most frequently 
carries positive parity, 
whereas for the heavy neutron-odd and light proton-odd fragments 
it more frequently carries negative parity.
These effects are less pronounced with the Skyrme than with the 
Gogny EDF, most likely because of the differences in predicted 
single-particle spectra. 
In particular, the enhanced population of negative-parity orbitals, 
such as the $1h_{11/2}$ shell illustrated in 
Fig.~\ref{fig:pu240_fragments}(g) for $^{133}$Sb,
increases the probability of negative-parity states 
for the heavy odd-proton states.
Finally, as a consequence of the unpaired neutron and proton having opposite parities, 
odd-odd fragments exhibit an enhanced population of negative-parity states, 
again more pronounced in the heavy fragments and
particularly in the Gogny case.

\textit{Conclusion.} 
Using simultaneous projections on the fragment
quantum numbers $J$, $K$, $\pi$, $N$, and $Z$, we have provided
a fully microscopic characterization of the spin--parity
content of fission fragments 
in the thermal neutron-induced
fission of $^{239}$Pu.
These predictions are consistent between two independent, 
state-of-the-art models, 
based on Gogny and Skyrme functionals.

The light, well-deformed fragment follows
a standard spin-cutoff behavior dominated
by collective effects, while the nearly spherical heavy
fragment is strongly influenced by shell structure. 
Dynamical pair
breaking is found to drive both the population of
unnatural-parity states and the generation
of non-zero
$K$ components, with high-$|K|$ configurations
tending toward an equal mixture of natural and
unnatural parity.
Parity distributions are also sensitive to the fragment's number parity, 
with odd-mass and odd-odd fragments showing 
distinct parity patterns tied to 
the orbitals of the unpaired nucleons.
These results show that the parity content of 
fission fragments at low
energy may depart significantly from the
equiprobable partition commonly assumed
in statistical de-excitation models.

The present study is based on a single
TDHFB trajectory corresponding to the most probable fission
path, and is therefore limited in its ability to capture
the full range of fission-fragment configurations.
Nonetheless, the microscopic mechanisms
governing the spin--parity distributions 
are expected to remain robust beyond the specific
trajectory considered here. A quantitative description
of the full range of fragments
will, however, require considering an ensemble of
fission trajectories, e.g. within the time-dependent
generator coordinate method \cite{verriere2020,marevic2023,li2023}.
Future work should also investigate the dependence of the 
spin-parity distributions on excitation energy, as well as how
incorporating them into de-excitation
simulations affects predictions
of observables
such as photon multiplicities, spectra,
and isomer population. Some of these directions have already been explored in a recent study \cite{marevic2026b}, which investigates the excitation-energy dependence of spin--parity distributions and their impact on prompt photon emission.

\textit{Acknowledgements.}
We gratefully acknowledge support from the CNRS/IN2P3 
supercomputer Center (Lyon, France) for providing 
computing and data-processing resources needed for this work. 
This work was also granted access to the HPC resources 
of IDRIS and CINES under Allocation No. 2024-AD010515531R1 made by GENCI.
The work of P.~M. was funded by the European Union’s
Horizon Europe research and innovation program under the
Marie Sk\l{}odowska-Curie Actions Grant Agreement No.~101149053.
Support for this work was partly provided through Scientific Discovery
through Advanced Computing (SciDAC) program funded by U.S. Department of
Energy, Office of Science, Advanced Scientific Computing Research and
Nuclear Physics. This work was partly performed under the auspices of 
the US Department of Energy by the Lawrence Livermore National Laboratory under 
Contract DE-AC52-07NA27344. Computing support for this work came from the Lawrence 
Livermore National Laboratory Institutional Computing Grand Challenge program.
G.S. acknowledges support from the French National Research Agency (ANR) under the Chaire de Professeur Junior programme (Grant No. ANR-23-CPJ1-0154).

\bibliography{bibliography}

@article{Kawano2023,
  title = {{Consideration of memory of spin and parity in the fissioning compound nucleus by applying the Hauser-Feshbach fission fragment decay model to photonuclear reactions}},
  author = {Kawano, T. and Lovell, A. E. and Okumura, S. and Sasaki, H. and Stetcu, I. and Talou, P.},
  journal = {Phys. Rev. C},
  volume = {107},
  issue = {4},
  pages = {044608},
  numpages = {9},
  year = {2023},
  month = {Apr},
  publisher = {American Physical Society},
  doi = {10.1103/PhysRevC.107.044608},
  url = {https://link.aps.org/doi/10.1103/PhysRevC.107.044608}
}

@article{wimmer12,
 author = {Wimmer, M.},
 title = {Algorithm 923: Efficient Numerical Computation of the Pfaffian for Dense and Banded Skew-Symmetric Matrices},
 journal = {ACM Trans. Math. Softw.},
 issue_date = {August 2012},
 volume = {38},
 number = {4},
 month = aug,
 year = {2012},
 issn = {0098-3500},
 pages = {30:1--30:17},
 articleno = {30},
 numpages = {17},
 url = {http://doi.acm.org/10.1145/2331130.2331138},
 doi = {10.1145/2331130.2331138},
 acmid = {2331138},
 publisher = {ACM},
 address = {New York, NY, USA},
}

@article{dobaczewski2009,
title = {{Solution of the Skyrme–Hartree–Fock–Bogolyubov equations in the Cartesian deformed harmonic-oscillator basis.: (VI) hfodd (v2.40h): A new version of the program}},
journal = {Computer Physics Communications},
volume = {180},
number = {11},
pages = {2361-2391},
year = {2009},
issn = {0010-4655},
doi = {https://doi.org/10.1016/j.cpc.2009.08.009},
url = {https://www.sciencedirect.com/science/article/pii/S0010465509002598},
author = {J. Dobaczewski and W. Satuła and B.G. Carlsson and J. Engel and P. Olbratowski and P. Powałowski and M. Sadziak and J. Sarich and N. Schunck and A. Staszczak and M. Stoitsov and M. Zalewski and H. Zduńczuk}
}

@article{li2023,
  title = {Generalized time-dependent generator coordinate method for small- and large-amplitude collective motion},
  author = {Li, B. and Vretenar, D. and Nik\ifmmode \check{s}\else \v{s}\fi{}i\ifmmode \acute{c}\else \'{c}\fi{}, T. and Zhao, P. W. and Meng, J.},
  journal = {Phys. Rev. C},
  volume = {108},
  issue = {1},
  pages = {014321},
  numpages = {12},
  year = {2023},
  month = {Jul},
  publisher = {American Physical Society},
  doi = {10.1103/PhysRevC.108.014321},
  url = {https://link.aps.org/doi/10.1103/PhysRevC.108.014321}
}

@article{marevic2023,
  title = {{Quantum fluctuations induce collective multiphonons in finite Fermi liquids}},
  author = {Marevi\ifmmode \acute{c}\else \'{c}\fi{}, Petar and Regnier, David and Lacroix, Denis},
  journal = {Phys. Rev. C},
  volume = {108},
  issue = {1},
  pages = {014620},
  numpages = {7},
  year = {2023},
  month = {Jul},
  publisher = {American Physical Society},
  doi = {10.1103/PhysRevC.108.014620},
  url = {https://link.aps.org/doi/10.1103/PhysRevC.108.014620}
}

@misc{marevic2026b,
  title = {Energy-Resolved Spin and Parity Distributions of Fission Fragments},
  author = {Petar Marević and Nicolas Schunck and Antonio Bjelčić and Guillaume Scamps},
  year = {2026},
  eprint = {2610.03018},
  archivePrefix = {arXiv},
  primaryClass = {nucl-th},
  url = {https://arxiv.org/abs/2610.03018}
}

@article{li2026,
  title = {Intrinsic Generation of Angular Momenta and Entanglement in Fission},
  author = {Li, B. and Zhang, D. D. and Vretenar, D. and Nik\ifmmode \check{s}\else \v{s}\fi{}i\ifmmode \acute{c}\else \'{c}\fi{}, T. and Zhao, P. W. and Meng, J.},
  journal = {Phys. Rev. Lett.},
  volume = {136},
  issue = {25},
  pages = {252502},
  numpages = {7},
  year = {2026},
  month = {Jun},
  publisher = {American Physical Society},
  doi = {10.1103/tyht-frnf},
  url = {https://link.aps.org/doi/10.1103/tyht-frnf}
}

@article{Marevic2026,
  title = {Microscopic theory of angular momentum distributions across the full range of fission fragments},
  author = {Marevi\ifmmode \acute{c}\else \'{c}\fi{}, Petar and Schunck, Nicolas and Verriere, Marc},
  journal = {Phys. Rev. C},
  volume = {113},
  issue = {1},
  pages = {014612},
  numpages = {18},
  year = {2026},
  month = {Jan},
  publisher = {American Physical Society},
  doi = {10.1103/yr2c-nvf3},
  url = {https://link.aps.org/doi/10.1103/yr2c-nvf3}
}

@article{schunck2022,
title = {Theory of nuclear fission},
journal = {Progress in Particle and Nuclear Physics},
volume = {125},
pages = {103963},
year = {2022},
issn = {0146-6410},
doi = {https://doi.org/10.1016/j.ppnp.2022.103963},
url = {https://www.sciencedirect.com/science/article/pii/S0146641022000242},
author = {Nicolas Schunck and David Regnier}
}

@article{bulgac2021,
  title = {Fission Fragment Intrinsic Spins and Their Correlations},
  author = {Bulgac, Aurel and Abdurrahman, Ibrahim and Jin, Shi and Godbey, Kyle and Schunck, Nicolas and Stetcu, Ionel},
  journal = {Phys. Rev. Lett.},
  volume = {126},
  issue = {14},
  pages = {142502},
  numpages = {7},
  year = {2021},
  month = {Apr},
  publisher = {American Physical Society},
  doi = {10.1103/PhysRevLett.126.142502},
  url = {https://link.aps.org/doi/10.1103/PhysRevLett.126.142502}
}

@article{marevic2021,
  title = {Angular momentum of fission fragments from microscopic theory},
  author = {Marevi\ifmmode \acute{c}\else \'{c}\fi{}, Petar and Schunck, Nicolas and Randrup, J\o{}rgen and Vogt, Ramona},
  journal = {Phys. Rev. C},
  volume = {104},
  issue = {2},
  pages = {L021601},
  numpages = {6},
  year = {2021},
  month = {Aug},
  publisher = {American Physical Society},
  doi = {10.1103/PhysRevC.104.L021601},
  url = {https://link.aps.org/doi/10.1103/PhysRevC.104.L021601}
}

@article{wilson2021,
  title = {{Angular momentum generation in nuclear fission}},
  author = {Wilson et al., J. N.},
  journal = {Nature},
  volume = {590},
  pages = {566-570},
  year = {2021},
  doi = {https://doi.org/10.1038/s41586-021-03304-w},
  url = {10.1038/s41586-021-03304-w}
}

@article{scamps2023b,
  title = {Spatial orientation of the fission fragment intrinsic spins and their correlations},
  author = {Scamps, Guillaume and Abdurrahman, Ibrahim and Kafker, Matthew and Bulgac, Aurel and Stetcu, Ionel},
  journal = {Phys. Rev. C},
  volume = {108},
  issue = {6},
  pages = {L061602},
  numpages = {6},
  year = {2023},
  month = {Dec},
  publisher = {American Physical Society},
  doi = {10.1103/PhysRevC.108.L061602},
  url = {https://link.aps.org/doi/10.1103/PhysRevC.108.L061602}
}

@book{talou2023,
  title     = {{Nuclear Fission: Theories, Experiments and Applications}},
  publisher = {Springer},
  year      = {2023},
  editor    = {Talou, P. and Vogt, R.}
}

@misc{SM,
  note = {See Supplemental Material at [URL will be inserted by publisher] 
  for more details on TDHFB and projected calculations, 
  a discussion of the microscopic connection between pair breaking 
  and the emergence of non-zero $K$ and unnatural-parity components, 
  and an assessment of the effect of global particle-number 
  projection on predictions
  of the model.}
}

@article{robledo2026,
  title = {Particle number projection on a spatial domain},
  author = {Robledo, L. M.},
  journal = {Phys. Rev. C},
  volume = {113},
  issue = {6},
  pages = {L061305},
  numpages = {6},
  year = {2026},
  month = {Jun},
  publisher = {American Physical Society},
  doi = {10.1103/mlnn-1y1h},
  url = {https://link.aps.org/doi/10.1103/mlnn-1y1h}
}

@article{robledo2009,
  title = {{Sign of the overlap of Hartree-Fock-Bogoliubov wave functions}},
  author = {Robledo, L. M.},
  journal = {Phys. Rev. C},
  volume = {79},
  issue = {2},
  pages = {021302(R)},
  numpages = {5},
  year = {2009},
  month = {Feb},
  publisher = {American Physical Society},
  doi = {10.1103/PhysRevC.79.021302},
  url = {https://link.aps.org/doi/10.1103/PhysRevC.79.021302}
}

@article{bertsch2012,
  title = {{Symmetry Restoration in Hartree-Fock-Bogoliubov Based Theories}},
  author = {Bertsch, G. F. and Robledo, L. M.},
  journal = {Phys. Rev. Lett.},
  volume = {108},
  issue = {4},
  pages = {042505},
  numpages = {4},
  year = {2012},
  month = {Jan},
  publisher = {American Physical Society},
  doi = {10.1103/PhysRevLett.108.042505},
  url = {https://link.aps.org/doi/10.1103/PhysRevLett.108.042505}
}

@Book{krappe2012,
  author    = {Krappe, H. J. and Pomorski, K.},
  publisher = {Springer},
  title     = {{Theory of Nuclear Fission}},
  year      = {2012},
}

@article{bulgac2022a,
  title = {Fragment Intrinsic Spins and Fragments' Relative Orbital Angular Momentum in Nuclear Fission},
  author = {Bulgac, Aurel and Abdurrahman, Ibrahim and Godbey, Kyle and Stetcu, Ionel},
  journal = {Phys. Rev. Lett.},
  volume = {128},
  issue = {2},
  pages = {022501},
  numpages = {6},
  year = {2022},
  month = {Jan},
  publisher = {American Physical Society},
  doi = {10.1103/PhysRevLett.128.022501},
  url = {https://link.aps.org/doi/10.1103/PhysRevLett.128.022501}
}

@incollection{Bulgac2019b,
  author    = {Bulgac, Aurel},
  title     = {Time-Dependent Density Functional Theory},
  booktitle = {Energy Density Functional Methods for Atomic Nuclei},
  editor    = {Schunck, Nicolas},
  publisher = {IOP Publishing},
  year      = {2019},
  chapter   = {4}
}

@article{simenel2025,
  title = {{Nuclear quantum many-body dynamics (2nd edition)}},
  author = {Simenel, Cedric},
  journal = {Eur. Phys. J. A},
  volume = {61},
  pages = {181},
  year = {2025},
  doi = {10.1140/epja/s10050-025-01624-3}
}

@article{hashimoto2012,
  title = {{Linear responses in time-dependent
Hartree-Fock-Bogoliubov method with Gogny interaction}},
  author = {Hashimoto, Y.},
  journal = {Eur. Phys. J. A},
  volume = {48},
  pages = {55},
  year = {2012},
  doi = {https://doi.org/10.1140/epja/i2012-12055-0}
}

@article{hashimoto2013,
  title = {{Time-dependent Hartree-Fock-Bogoliubov calculations using a Lagrange mesh with the Gogny interaction}},
  author = {Hashimoto, Yukio},
  journal = {Phys. Rev. C},
  volume = {88},
  issue = {3},
  pages = {034307},
  numpages = {7},
  year = {2013},
  month = {Sep},
  publisher = {American Physical Society},
  doi = {10.1103/PhysRevC.88.034307},
  url = {https://link.aps.org/doi/10.1103/PhysRevC.88.034307}
}

@article{hashimoto2016,
  title = {{Gauge angle dependence in time-dependent Hartree-Fock-Bogoliubov calculations of $^{20}\mathrm{O}+^{20}\mathrm{O}$ head-on collisions with the Gogny interaction}},
  author = {Hashimoto, Yukio and Scamps, Guillaume},
  journal = {Phys. Rev. C},
  volume = {94},
  issue = {1},
  pages = {014610},
  numpages = {10},
  year = {2016},
  month = {Jul},
  publisher = {American Physical Society},
  doi = {10.1103/PhysRevC.94.014610},
  url = {https://link.aps.org/doi/10.1103/PhysRevC.94.014610}
}

@article{scamps2017,
  title = {{Transfer probabilities for the reactions $^{14,20}\mathrm{O}+^{20}\mathrm{O}$ in terms of multiple time-dependent Hartree-Fock-Bogoliubov trajectories}},
  author = {Scamps, Guillaume and Hashimoto, Yukio},
  journal = {Phys. Rev. C},
  volume = {96},
  issue = {3},
  pages = {031602(R)},
  numpages = {6},
  year = {2017},
  month = {Sep},
  publisher = {American Physical Society},
  doi = {10.1103/PhysRevC.96.031602},
  url = {https://link.aps.org/doi/10.1103/PhysRevC.96.031602}
}

@article{berger1984,
title = {Microscopic analysis of collective dynamics in low energy fission},
journal = {Nuclear Physics A},
volume = {428},
pages = {23-36},
year = {1984},
issn = {0375-9474},
doi = {https://doi.org/10.1016/0375-9474(84)90240-9},
url = {https://www.sciencedirect.com/science/article/pii/0375947484902409},
author = {J.F. Berger and M. Girod and D. Gogny}
}

@article{bjelcic2026,
  title = {Excitation energy of fission fragments within nuclear time-dependent density functional theory},
  author = {Bjel\ifmmode \check{c}\else \v{c}\fi{}i\ifmmode \acute{c}\else \'{c}\fi{}, Antonio and Schunck, Nicolas and Verriere, Marc},
  journal = {Phys. Rev. C},
  volume = {113},
  issue = {3},
  pages = {034602},
  numpages = {24},
  year = {2026},
  month = {Mar},
  publisher = {American Physical Society},
  doi = {10.1103/tgt3-zwn7},
  url = {https://link.aps.org/doi/10.1103/tgt3-zwn7}
}

@misc{scamps2025,
      title={Uncertainty Principle and Angular Momentum Generation in Microscopic Fission Models}, 
      author={G. Scamps and A. Guilleux and D. Regnier and A. Bernard},
      year={2025},
      eprint={2512.02207},
      archivePrefix={arXiv},
      primaryClass={nucl-th},
      url={https://arxiv.org/abs/2512.02207}, 
}

@article{verbeke2018,
title = "{Fission Reaction Event Yield Algorithm FREYA 2.0.2}",
journal = "Computer Physics Communications",
volume = "222",
pages = "263 - 266",
year = "2018",
issn = "0010-4655",
doi = "https://doi.org/10.1016/j.cpc.2017.09.006",
url = "http://www.sciencedirect.com/science/article/pii/S001046551730293X",
author = "J. M. Verbeke and J. Randrup and R. Vogt"
}

@article{talou2021,
title = {Fission fragment decay simulations with the {CGMF} code},
journal = {Computer Physics Communications},
volume = {269},
pages = {108087},
year = {2021},
issn = {0010-4655},
doi = {https://doi.org/10.1016/j.cpc.2021.108087},
url = {https://www.sciencedirect.com/science/article/pii/S0010465521001995},
author = {P. Talou and I. Stetcu and P. Jaffke and M.E. Rising and A.E. Lovell and T. Kawano}
}

@article{litaize2015,
  title = {Fission modelling with {FIFRELIN}},
  author = {Litaize, O. and Serot, O. and Berge, L.},
  journal = {Eur. Phys. J. A},
  volume = {51},
  pages = {177},
  year = {2015},
  doi = {10.1140/epja/i2015-15177-9},
  url = {https://link.springer.com/article/10.1140/epja/i2015-15177-9#citeas}
}

@Article{bartel1982,
  author  = {Bartel, J. and Quentin, P. and Brack, M. and Guet, C. and H{\aa}kansson, H.-B.},
  journal = {Nucl. Phys. A},
  title   = {Towards a better parametrisation of {{Skyrme}}-like effective forces: {A} critical study of the {{SkM}} force},
  year    = {1982},
  number  = {1},
  pages   = {79},
  volume  = {386},
  doi     = {10.1016/0375-9474(82)90403-1},
}

@InCollection{dobaczewski2002,
  author    = {Dobaczewski, J. and Nazarewicz, W. and Stoitsov, M. V.},
  booktitle = {The {{Nuclear Many}}-{{Body Problem}} 2001},
  publisher = {{Springer Netherlands}},
  title     = {{Contact Pairing Interaction for the {{Hartree}}-{{Fock}}-{{Bogoliubov}} Calculations}},
  year      = {2002},
  isbn      = {978-94-010-0460-2},
  number    = {53},
  pages     = {181},
  series    = {Nato {{Science Series II}}},
}

@article{scamps2013,
  title = {{Effect of pairing on one- and two-nucleon transfer below the Coulomb barrier: A time-dependent microscopic description}},
  author = {Scamps, Guillaume and Lacroix, Denis},
  journal = {Phys. Rev. C},
  volume = {87},
  issue = {1},
  pages = {014605},
  numpages = {11},
  year = {2013},
  month = {Jan},
  publisher = {American Physical Society},
  doi = {10.1103/PhysRevC.87.014605},
  url = {https://link.aps.org/doi/10.1103/PhysRevC.87.014605}
}

@article{simenel2010,
  title = {{Particle Transfer Reactions with the Time-Dependent Hartree-Fock Theory Using a Particle Number Projection Technique}},
  author = {Simenel, C\'edric},
  journal = {Phys. Rev. Lett.},
  volume = {105},
  issue = {19},
  pages = {192701},
  numpages = {4},
  year = {2010},
  month = {Nov},
  publisher = {American Physical Society},
  doi = {10.1103/PhysRevLett.105.192701},
  url = {https://link.aps.org/doi/10.1103/PhysRevLett.105.192701}
}

@article{bertsch2019,
  title = {Angular momentum of fission fragments},
  author = {Bertsch, G. F. and Kawano, T. and Robledo, L. M.},
  journal = {Phys. Rev. C},
  volume = {99},
  issue = {3},
  pages = {034603},
  numpages = {5},
  year = {2019},
  month = {Mar},
  publisher = {American Physical Society},
  doi = {10.1103/PhysRevC.99.034603},
  url = {https://link.aps.org/doi/10.1103/PhysRevC.99.034603}
}

@misc{bohr1970nuclear,
  title={Nuclear structure, vol. 1},
  author={Bohr, Aage and Mottelson, Ben R and Breit, Gregory},
  year={1970},
  publisher={American Institute of Physics}
}

@article{verriere2020,
  title = {The {{Time-Dependent Generator Coordinate Method}} in {{Nuclear Physics}}},
  author = {Verriere, Marc and Regnier, David},
  year = {2020},
  journal = {Front. Phys.},
  volume = {8},
  pages = {1},
  publisher = {Frontiers},
  doi = {10.3389/fphy.2020.00233},
  langid = {english}
}

@Book{ring2004,
  author    = {Ring, P. and Schuck, P.},
  publisher = {Springer},
  title     = {{The Nuclear Many-Body Problem}},
  year      = {2004},
  series    = {Texts and Monographs in Physics},
}

\end{document}


\title{Supplemental Material:\\
Microscopic Spin--Parity Distributions of Fission Fragments}

\author{Guillaume Scamps}
\affiliation{Universit\'e de Toulouse, CNRS/IN2P3, L2IT, Toulouse, France}

\author{Petar Marevi\'c}
\affiliation{University of Zagreb, Faculty of Science, Department of Physics,
HR-10000 Zagreb, Croatia}

\author{Antonio Bjel\v{c}i\'c}
\affiliation{Nuclear data and Theory Group, Nuclear and 
Chemical Science Division, Lawrence Livermore National Laboratory, 
California, USA 94550}

\author{Nicolas Schunck}
\affiliation{Nuclear data and Theory Group, Nuclear 
and Chemical Science Division, Lawrence Livermore National Laboratory, 
California, USA 94550}

\date{\today}

\maketitle


\section{Additional Details of the Projection Method}

\subsection{Parity operator}

The full joint distribution of the fission fragments' 
angular momentum $J$, its projection
on the fission axis $K$, the parity $\pi$, the neutron number $N$,
and the proton number $Z$ is computed as
\begin{widetext}
\begin{align}
P_{F}(J,K,\pi,N,Z)
 = \frac{2J+1}{128\pi^4}  &  \int_0^{2\pi} d\varphi_n  
 \int_0^{2\pi} d\varphi_p \int_0^{2\pi} d\alpha \int_0^{\pi} d\beta 
 \, \sin\beta \int_0^{4\pi} d\gamma \, 
 {\cal D}^{J*}_{K K} (\alpha, \beta, \gamma) \, \nonumber \\
& \times \langle \Psi | \left( \hat{1} + \pi \hat{\Pi}_F \right) 
e^{i \varphi_p (\hat{Z}_F-Z)}  e^{i \varphi_n (\hat{N}_F-N)}  
e^{i\alpha \hat J_{zF}} e^{i\beta \hat J_{yF}} 
e^{i\gamma \hat J_{zF}} | \Psi \rangle.
\label{eq:PJMKpiNZ}
\end{align}
\end{widetext}

The angular momentum operators in the fragments
are defined equivalently to Ref.~\cite{scamps2023b},
and the particle number operators
equivalently to Ref.~\cite{marevic2026}.
Prior to the angular momentum projection,
each fragment is brought to its rest frame
like in previous TDHFB studies 
\cite{bulgac2021,bulgac2022a,scamps2023b,scamps2025,li2026}.
The local space reflection operator is defined as
\begin{equation}
\hat{\Pi}_F = (1-\hat{\Theta}_F) + 
\hat{T}_F^\dagger\hat{\Theta}_F\hat{\Pi}\hat{\Theta}_F\hat{T}_F.
\end{equation}
Here, $\hat{\Theta}_F$ is the characteristic operator for the 
spatial region occupied by fragment $F$, equal to $1$ inside and
$0$ outside; $\hat{\Pi}$ is the 
standard space-inversion operator about the 
global coordinate origin, 
and $\hat{T}_F$ is the translation operator ensuring that
the inversion is performed in the fragment's own frame.

\subsection{Gogny calculations}

Gogny-TDHFB calculations are identical to those
in Ref.~\cite{scamps2025} for $^{240}$Pu.
The basis consists of two-dimensional harmonic oscillator states 
restricted to $n_x + n_y \leq N_{\mathrm{shell}}$ with
$N_{\mathrm{shell}} = 9$ and a one-dimensional mesh along the $z$ axis.
The number of quasiparticle states is entirely fixed by
the basis size, with $N_{\mathrm{qp}} = N_{\mathrm{basis}} = 5720$.
No spatial symmetry is imposed and the time-reversal
symmetry is broken during the TDHFB evolution.

A Bloch-Messiah decomposition 
of the TDHFB wave packet is performed
for the well-separated fragments. 
The single particle wave functions in the canonical basis 
are then projected onto a Cartesian grid of 
size $30 \times 30 \times 52$ with a mesh spacing 
of $\Delta x=0.8$ fm.  
The overlap between the original state and the 
rotated state is computed with the Pfaffian method using 
the algorithm of Ref.~\cite{wimmer12}.

The  $\alpha$ and $\gamma$ integrals in Eq.~\eqref{eq:PJMKpiNZ} are 
discretized with a uniform spacing of $\pi/8$.
The angle $\beta$ is discretized using a 
Gauss-Legendre quadrature in $\beta$ 
(as in Ref.~\cite{dobaczewski2009}). 
In the present calculations, 
the integral over $\beta$ is approximated with $N_\beta = 32$ points.
Using symmetries, the gauge angle integrations are 
reduced to a range from $[0,\pi]$. The gauge angle rotations
over this interval are performed with 9 points. 

\begin{figure}
    \centering
\includegraphics[width=0.85\linewidth]{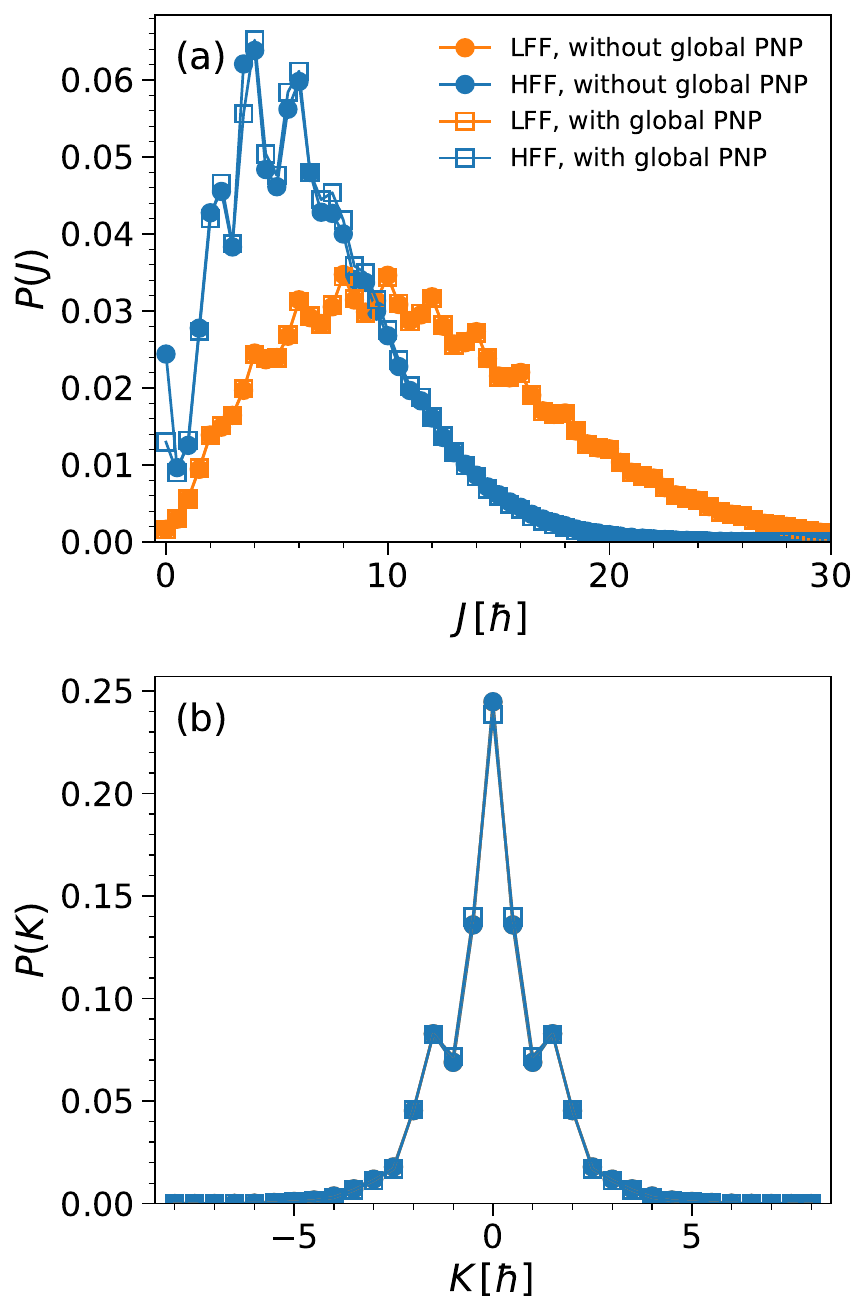}
   \caption{Distributions of the fission-fragment spin $J$ (a)
   and its fission-axis projection $K$ (b), obtained by marginalizing the full distribution, with and without
   the global PNP. Results are shown
   for  both fragments, except in panel
   (b), where the LFF and HFF
   distributions are identical and only
   the HFF is shown.}
   \label{fig:supp_globalPNP_JK}
\end{figure}

\begin{figure}
    \centering
\includegraphics[width=0.85\linewidth]{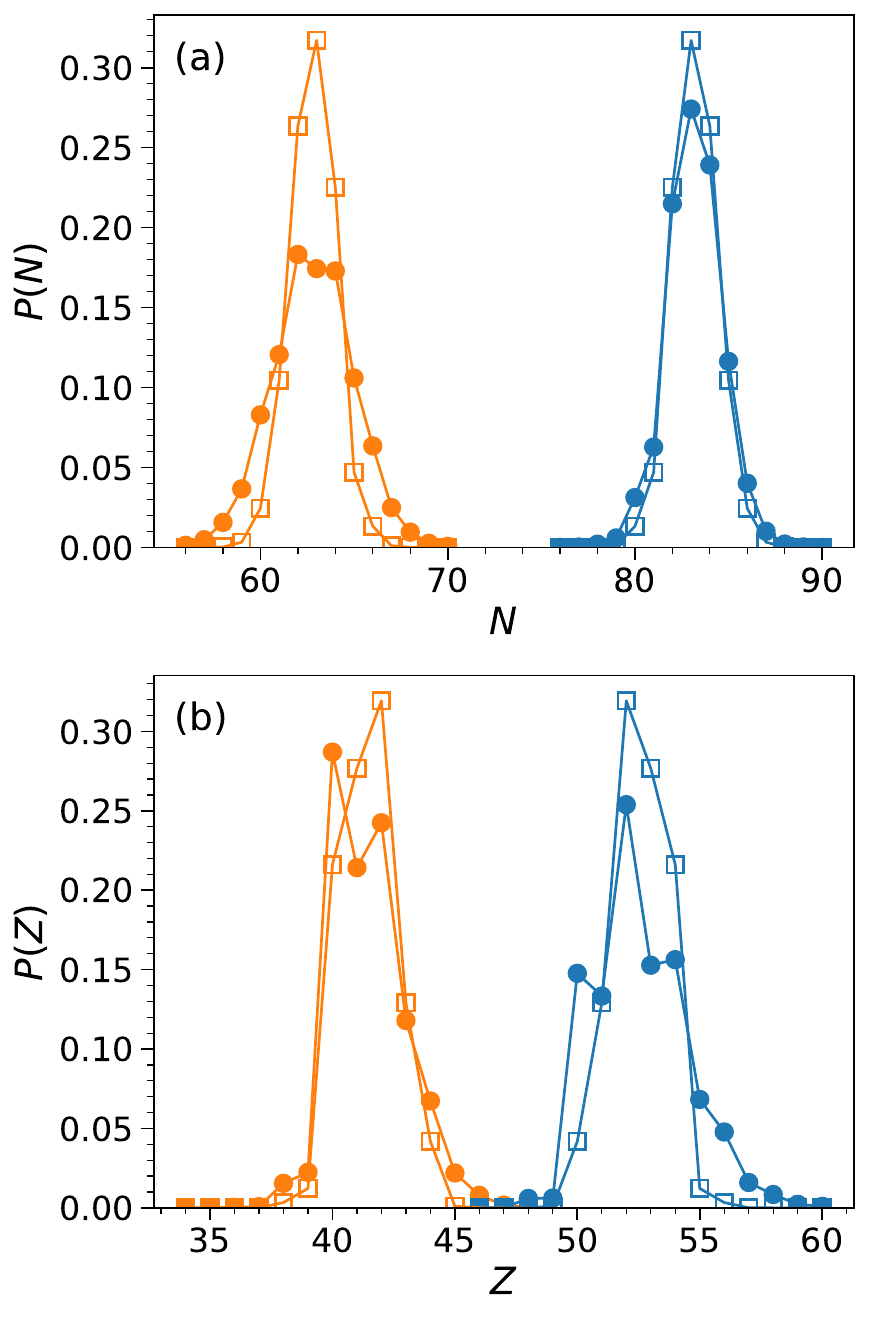}
   \caption{The same as Fig.~\ref{fig:supp_globalPNP_JK},
   but for distributions of the fission-fragment neutron number (a)
   and proton number (b).
    Colors and markers follow the same convention 
    as in Fig.~\ref{fig:supp_globalPNP_JK}.}
   \label{fig:supp_globalPNP_NZ}
\end{figure}

\subsection{Skyrme calculations}

Skyrme-TDHFB calculations are performed with the newly developed
\axialhohfb code \cite{bjelcic2026}. Single-nucleon wave functions are expanded
in the basis of the axially-symmetric harmonic
oscillator. This basis is uniquely defined
by three parameters: $Z_{\mathrm{box}}, R_{\mathrm{box}}$, and
$\max\{n_z\}$. Here, $2 Z_{\mathrm{box}}$ is the height
of the simulation box cylinder, $2 R_{\mathrm{box}}$
is its diameter, and $\max\{n_z\}$ is the largest
value of the $n_z$ quantum number present
in the basis. More details on the definition of the
basis
are available in Ref.~\cite{bjelcic2026}.

In this work, we set $2 Z_{\mathrm{box}} = 60$~fm,
$2 R_{\mathrm{box}} = 25$~fm and $\max\{n_z\} = 38$,
both for static and dynamic calculations.
This amounts to a total of $5967$ basis states.
Such a basis provides a very good convergence
profile and allows for the accommodation
of very elongated post-fission shapes.

The initial state is determined through
constrained HFB 
calculations, with constraints
imposed on values of the quadrupole deformation
parameter $q_{20}$ and the octupole
deformation parameter $q_{30}$. We choose an HFB configuration 
($q_{20} = 171.6$~b, $q_{30} = 13.2$~b$^{3/2}$) beyond the saddle
point. This configuration is evolved
in time using a fourth-order Runge-Kutta 
method with time step $\Delta t = 0.5$ fm/c. 
The time evolution is stopped once the distance
between the fragments' centers of mass
exceeds $30$~fm. The resulting heavy fragment has,
on average, $\braket{N_H} = 83.18$ and 
$\braket{Z_H} = 52.56$, while the light fragment has
$\braket{N_L} = 62.82$ and $\braket{Z_L} = 41.44$.

Projections are performed in the canonical basis \cite{scamps2013}, using the Pfaffian
method \cite{bertsch2012,wimmer12}.
A uniform mesh of $N_{\alpha}=N_{\gamma} = 18$ points 
is used for Euler angles
$\alpha$ and $\gamma$ in Eq.~\eqref{eq:PJMKpiNZ},
while a Gauss-Legendre quadrature
with $N_{\beta} = 42$ points is used for the Euler angle $\beta$.
Particle number projection is performed for both isospins with an equidistant
mesh of $N_\varphi = 17$ points in the $[0, 2\pi]$ interval.

\section{The effect of global particle number projection}

In Eq.~\eqref{eq:PJMKpiNZ} we perform the particle number
projection (PNP) in the fragments only. An additional global PNP could be 
performed, such as the one in Ref.~\cite{marevic2026},
that ensures the correct number of particles in the total
system. Such a procedure is expected to have a substantial
impact on the distributions of neutron and proton
numbers in the fragments and, consequently, fission yields.
However, a much smaller impact is expected for the distributions
of spin and its
fission-axis component. In this work, 
Skyrme calculations were performed both
with and without the global PNP.
In order to keep a direct comparison
with the Gogny results
--- which were
obtained without the global 
PNP --- the main text
presents Skyrme results without global PNP.
In this section, we briefly quantify the
consequences of this choice.

In Fig.~\ref{fig:supp_globalPNP_JK} we 
show the marginalized $P(J)$ and $P(K)$
distributions, obtained
with and without the global PNP.
The difference in $P(J)$ is negligible 
at higher $J$; at lower $J$,
the HFF distribution shows small
point-to-point differences between
the two treatments, most notably
a reduction of $P(J=0)$
with global PNP.
Furthermore,
the $P(K)$ distribution is essentially unchanged;
here we show it only for the HFF since the
LFF distribution is equal due to the 
$K_{\mathrm{tot}} = 0$ restriction for the full system.
Fig.~\ref{fig:supp_globalPNP_NZ} shows 
the distributions in $P(N)$ and $P(Z)$.
As expected, these distributions
are affected by the global PNP.
Most notably, by enforcing the correct 
total number of particles,
the probability of finding $Z$ protons 
in one fragment must equal the probability
of finding $Z_0 - Z$ protons in the
complementary fragment (and likewise
for $N$), where $Z_0$ ($N_0$) is the
proton (neutron) number of the total system. Consequently,
the models aimed at predicting
fission yields must include the
global PNP.

\begin{figure}
    \centering
\includegraphics[width=0.95\linewidth]{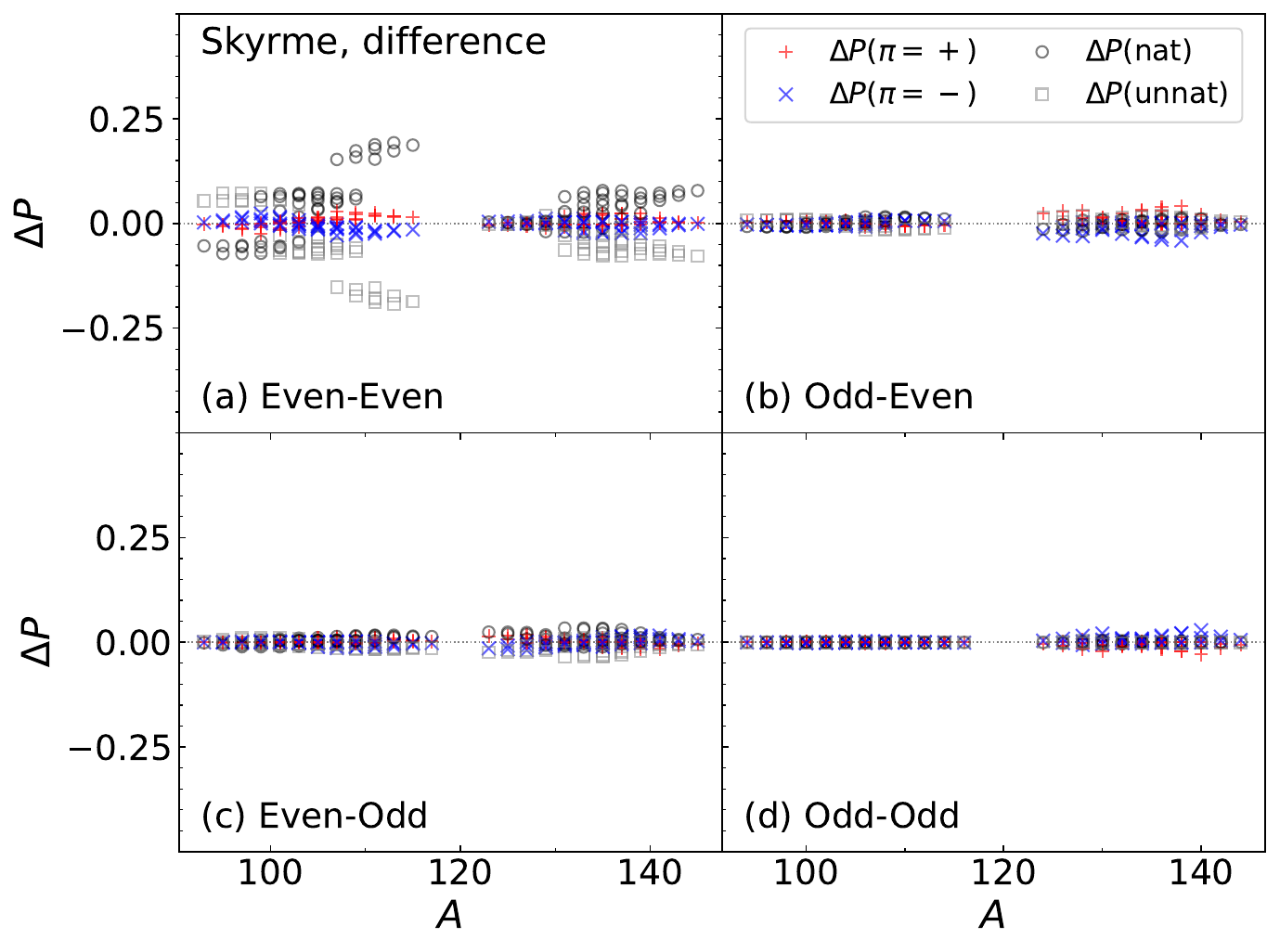}
   \caption{Difference $\Delta P = P_{\mathrm{with}} - P_{\mathrm{without}}$ 
   between parity distributions obtained with and without
   the global PNP, as a function of fragment mass
   number $A$, partitioned as in Fig.~3 of the
   main text: Even-Even (a), Odd-Even (b),
   Even-Odd (c), and Odd-Odd (d).
   Shown are the differences in the positive-parity
   (red pluses), negative-parity (blue crosses),
   natural-parity (black circles),
   and unnatural-parity (gray squares)
   fractions.}
   \label{fig:supp_Fig3_delta}
\end{figure}

To assess the impact of the global
PNP on the predictions presented
in the manuscript, in Fig.~\ref{fig:supp_Fig3_delta}
we show the difference
in the parity distribution
as a function of fragment mass number,
obtained with and without the global 
PNP. As in Fig.~3 of the main text,
the fragments are partitioned
into four categories. 
The
largest discrepancy between
the two schemes for the positive/negative parity
is rather small and
is found for the ($N=82$, $Z=57$) fragment, $\Delta P(\pi=+) = 0.04$.
The natural parity distributions
are affected to a larger
extent, with the largest deviation
of $\Delta P(\mathrm{nat}) = 0.19$ for the
$(N=66, Z = 48)$ fragment.
These larger
discrepancies are found for fragments
populating the low-probability
tails of $P(N)$ and/or $P(Z)$
(Fig.~\ref{fig:supp_globalPNP_NZ}).
Even with these modifications,
however, the parity distribution
retains the same qualitative
character under both projection
schemes -- e.g., natural parity remains dominant
in even-even fragments.
The conclusions of this work are therefore 
unaffected by the choice of
projection scheme, both because
the fragments most sensitive
to this choice carry low statistical
weight and because
their qualitative parity
content is preserved.


\section{Pair breaking and fragment spin--parity content in a BCS state}
\label{app:pair_breaking}


The generation of angular momentum in fission fragments caused by the pair-breaking mechanism is most effectively illustrated by using the BCS form of the total wavefunction
\begin{equation}
 |\mathrm{HFB}\rangle =
 \prod_{\mu>0}
 \left(
 u_\mu+v_\mu
 a^\dagger_\mu a^\dagger_{\bar\mu}
 \right)|0\rangle ,
\end{equation}
where $\mu$ and $\bar\mu$ denote time-reversed single-particle states. For an axially symmetric system,
\begin{equation}
 K_{\bar\mu}=-K_\mu ,
\end{equation}
so that an intact Cooper pair carries
\begin{equation}
 K_{\mathrm{pair}}=K_\mu+K_{\bar\mu}=0.
\end{equation}

During the separation of the fragments, a single-particle state can in general have components on both sides. We write
\begin{equation}
 a^\dagger_\mu =
 c_{\mu L}a^\dagger_{\mu L}
 +c_{\mu R}a^\dagger_{\mu R},
\end{equation}
and similarly
\begin{equation}
 a^\dagger_{\bar\mu} =
 c_{\bar\mu L}a^\dagger_{\bar\mu L}
 +c_{\bar\mu R}a^\dagger_{\bar\mu R}.
\end{equation}
The states $a^\dagger_{\mu L}$ and $a^\dagger_{\mu R}$ denote the components associated with the left and right fragments, respectively\footnote{Note that, strictly speaking, $a^{\dagger}_{\mu} = \sum_{\alpha}\big( L_{\mu\alpha} a^{\dagger}_{\alpha R} + P_{\mu\alpha} a^{\dagger}_{\alpha L} \big)$ (similarly for $a_{\mu}$) in order for the $a^{\dagger}_{R},a_{R}$ and $a^{\dagger}_{L},a_{L}$ to obey the standard anticommutation relation of Fermion operators. This does not change the qualitative picture of our argument.}. Complete localization in one fragment corresponds to the limiting cases
\begin{equation}
 a^\dagger_\mu=a^\dagger_{\mu L}
 \qquad\text{or}\qquad
 a^\dagger_\mu=a^\dagger_{\mu R},
\end{equation}
i.e. $|c_{\mu L}|=1$ or $|c_{\mu R}|=1$. This case correspond to an unbroken pair.

Using the left--right decomposition, each factor of the BCS wave function can be written as
\begin{align}
 u_\mu+v_\mu & a^\dagger_\mu a^\dagger_{\bar\mu}
 = \nonumber\\ u_\mu
 &+v_\mu c_{\mu L}c_{\bar\mu L}
 a^\dagger_{\mu L}a^\dagger_{\bar\mu L}  
 +v_\mu c_{\mu R}c_{\bar\mu R}
 a^\dagger_{\mu R}a^\dagger_{\bar\mu R}
 \nonumber\\
 &+v_\mu c_{\mu L}c_{\bar\mu R}
 a^\dagger_{\mu L}a^\dagger_{\bar\mu R}
 +v_\mu c_{\mu R}c_{\bar\mu L}
 a^\dagger_{\mu R}a^\dagger_{\bar\mu L}.
 \label{eq:BCS}
\end{align}

\subsection{Pair breaking and fragment $K$}

The first two terms with pair creation operators in Eq.\eqref{eq:BCS} correspond to the two members of the pair being associated with the same fragment. In these configurations their angular-momentum projections cancel. For example, for the term $a^\dagger_{\mu L}a^\dagger_{\bar\mu L}$, we get
\begin{equation}
K_L = K_\mu+K_{\bar\mu}=0,
\end{equation}
and the pair therefore gives no intrinsic $K$ to that fragment.

The last two terms are qualitatively different: the two members of the time-reversed pair are distributed between the two fragments. For example, the component
\begin{equation}
 a^\dagger_{\mu L}a^\dagger_{\bar\mu R}
\end{equation}
has
\begin{equation}
 K_L=K_\mu\qquad
 K_R=K_{\bar\mu}=-K_\mu.
\end{equation}
Thus
\begin{equation}
 K_L=-K_R\neq0,
 \qquad
 K_{\mathrm{tot}}=K_L+K_R=0.
\end{equation}
Therefore, each fragment has a non-zero intrinsic $K$ ($K_L$ and $K_R$, respectively) but the full nucleus has $K=K_L+K_R=0$ has expected.
The same argument applies to the component
$a^\dagger_{\mu R}a^\dagger_{\bar\mu L}$.

This provides a simple microscopic picture of the generation of non-zero fragment $K$. As long as the two time-reversed nucleons remain associated with the same fragment, they contribute as an intact pair and their projections cancel locally. When the pair is broken and its two members are distributed between the fragments, each fragment can carry a non-zero projection, while the total projection remains exactly zero.

In this sense, pair breaking converts the correlation
\begin{equation}
 (K_\mu,-K_\mu)
\end{equation}
of an intact Cooper pair into correlated fragment angular momenta,
\begin{equation}
 K_L=-K_R.
\end{equation}

This provides a simple microscopic picture of the generation of non-zero fragment $K$. 
A single broken pair can distribute its two nucleons between the two fragments, but this
configuration leaves one unpaired nucleon in each fragment and therefore corresponds to
an odd--odd fragmentation. To obtain even fragments with non-zero $K$, two pairs must
be broken, so that two unpaired nucleons are present in each fragment.

Consider two initially paired configurations, $(\mu_1,\bar\mu_1)$ and
$(\mu_2,\bar\mu_2)$, with
\begin{equation}
 K_{\bar\mu_i}=-K_{\mu_i},
 \qquad i=1,2.
\end{equation}
A component in which the two members of each broken pair are distributed between the
fragments can then be written as
\begin{equation}
 a^\dagger_{\mu_1 L}
 a^\dagger_{\mu_2 L}
 a^\dagger_{\bar\mu_1 R}
 a^\dagger_{\bar\mu_2 R}.
\end{equation}
The two fragments have now an even number of particles, while their angular-momentum projections are
\begin{equation}
 K_L=K_{\mu_1}+K_{\mu_2},
 \qquad
 K_R=K_{\bar\mu_1}+K_{\bar\mu_2}
     =-K_{\mu_1}-K_{\mu_2},
\end{equation}
that still satisfy
\begin{equation}
 K_L=-K_R\neq0,
 \qquad
 K_{\mathrm{tot}}=K_L+K_R=0.
\end{equation}
Thus, two-pair breaking provides a direct microscopic mechanism for generating
non-zero $K$ in even fragments while preserving the exact vanishing of the total
intrinsic projection.

\subsection{Relation to the spin--parity distribution}

The same pair-breaking mechanism also enlarges the set of spin--parity
configurations accessible to an even fragment. 
If a fragment were a fully paired core, its collective rotational
configurations would have the usual natural-parity relation between the parity $\pi$ and angular momentum $J$,
\begin{equation}
 \pi=(-1)^J .
\end{equation}
Breaking two pairs then adds two unpaired nucleons to this paired core.
These nucleons can occupy single-particle configurations around the Fermi energy characterized
by arbitrary angular momenta $j_1$ and $j_2$ and parities $\pi_1$ and
$\pi_2$. Their angular momenta can couple to a range of total angular
momenta,
\begin{equation}
 |j_1-j_2| \leq j_{12} \leq j_1+j_2 ,
\end{equation}
while their parity is determined by the intrinsic configurations of the
two unpaired nucleons. Consequently, the resulting fragment state is no
longer constrained to the natural-parity relation $\pi=(-1)^J$.

Pair breaking therefore provides a mechanism for populating
unnatural-parity states in even fragments, in addition to the
configurations already present in the paired core. It does not imply
that every pair-breaking configuration has unnatural parity; rather,
breaking the pairs removes the restrictions imposed by the paired
vacuum and opens access to a larger spin--parity space.

This mechanism is distinct from static deformation effects. Octupole
deformation can generate negative-parity components while preserving
axial symmetry, whereas triaxial deformation breaks axial symmetry and
can generate intrinsic $K$-mixing. In the present calculations, the
fragment shapes remain axially symmetric. Static triaxiality is therefore
excluded as the origin of the observed non-zero $K$ components, which can
instead be associated with the dynamical breaking of pairing correlations
during the fission process.


\bibliography{bibliography}